\documentclass[runningheads]{llncs}

\usepackage{eccv}

\usepackage{eccvabbrv}

\usepackage{graphicx}
\usepackage{booktabs}

\usepackage[accsupp]{axessibility}  

\usepackage{xcolor}
\usepackage{colortbl}
\usepackage{multirow}
\definecolor{dr1}{rgb}{0.80, 0.6, 1.0}
\definecolor{Second}{rgb}{0.97,0.81,0.63}
\definecolor{dr2}{rgb}{0.9, 0.8, 1.0}
\definecolor{dr3}{rgb}{0.96, 0.92, 1.0}
\definecolor{x3}{rgb}{0.906, 0.98, 1.0}
\definecolor{x1}{rgb}{0.52, 0.92, 1.0}
\definecolor{x2}{rgb}{0.8, 0.97, 1.0}

\usepackage{hyperref}

\usepackage{orcidlink}
\usepackage{marvosym}
\usepackage{wrapfig}

\begin{document}

\title{DANTE-W: Diffuse Albedo Neural Texturing \protect\\ in the Wild}

\titlerunning{DANTE-W}

\author{Guangyu Wang\inst{1}\textsuperscript{$\star$}\orcidlink{0009-0000-5674-2642} \and
Tianheng Lu\inst{1}\textsuperscript{$\star$}\orcidlink{0009-0007-7163-5539} \and
Ruqi Huang\inst{1}\textsuperscript{\Letter}\orcidlink{0000-0001-5942-3671} \and Lu Fang\inst{1}\textsuperscript{\Letter}\orcidlink{0000-0003-3552-0367}}

\authorrunning{G.~Wang et al.}

\institute{\textsuperscript{1}Tsinghua University, Beijing 100084, China} 

\maketitle
\footnotetext[1]{\textsuperscript{\Letter}Correspondence to: Lu Fang (fanglu@tsinghua.edu.cn, \href{http://www.luvision.net/}{luvision.net}), Ruqi Huang (ruqihuang@sz.tsinghua.edu.cn, \href{https://rqhuang88.github.io/}{rqhuang88.github.io}).}
\footnotetext[2]{\textsuperscript{$\star$}Authors contributed equally to this work.}
\begin{abstract}
Classical mesh texturing techniques blend captured multi-view images directly, which inevitably suffer from baked-in shading and casted shadows that compromise visual fidelity during relighting. To circumvent this issue, we present a neural texturing framework, namely \textsc{Dante-w}, to enable high-fidelity diffuse albedo texture recovery from unstructured image collections for large-scale, in-the-wild scenes, which integrates seamlessly with traditional 3D reconstruction pipelines. Given a reconstructed mesh and its surface parameterization, our method fuses view-space generative albedo priors into a coherent texture space via an expressive neural representation, while substantially enhancing fine-grained textural details through physically principled neural rendering. 
To comprehensively evaluate our method, we curate a benchmark dataset featuring diverse, fine-grained textures, comprising both real-world in-the-wild scenes and synthetic objects. Extensive experiments verify the effectiveness of our approach in reconstructing accurate albedo textures and boosting relighting fidelity. Project page: \href{https://dante-wild.github.io/}{dante-wild.github.io}.
\keywords{Texture mapping \and Neural representation \and Diffuse albedo estimation}
\vspace{-1.0em}
\begin{figure*}[htbp]
    \centering
    \newcommand{\colw}{0.19}
    \newcommand{\figw}{1} 
    \includegraphics[width=\figw\textwidth,trim={0cm 0cm 0cm 0cm},clip]{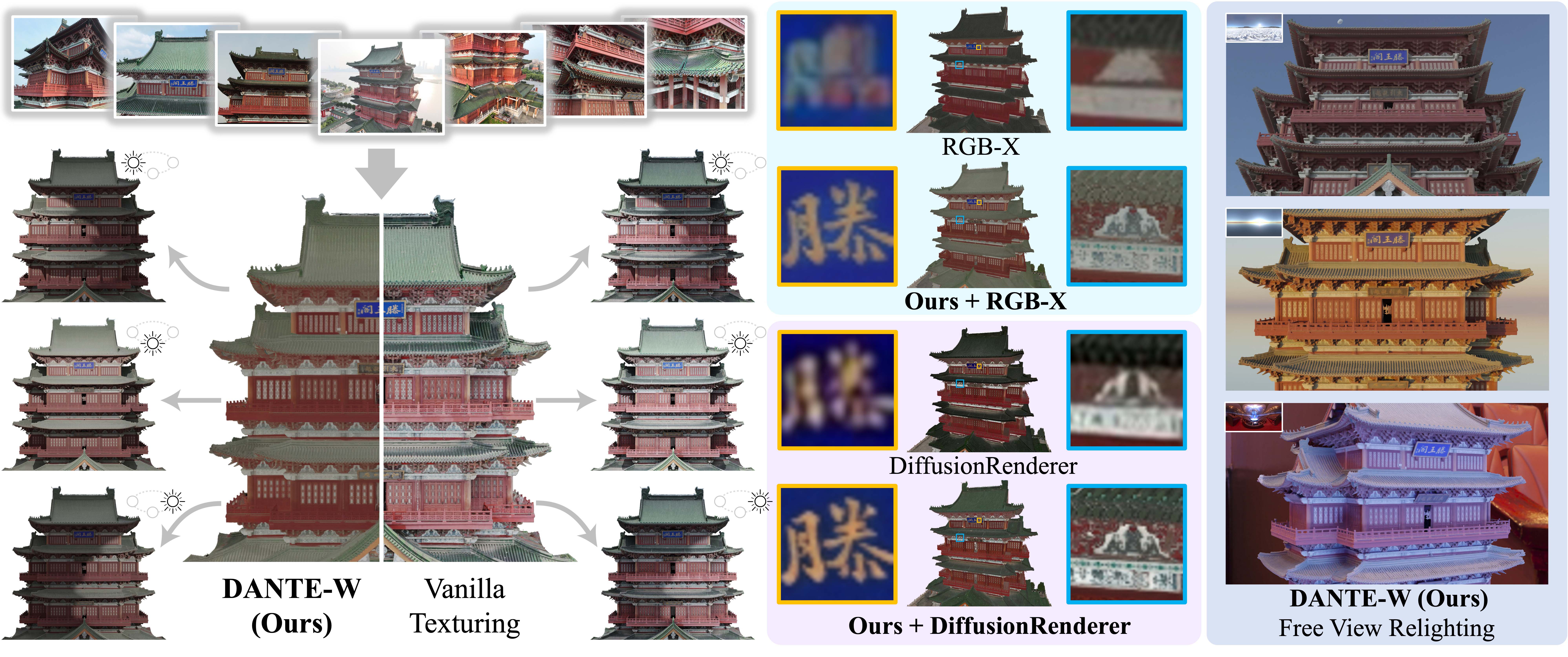}
    \hfill
\vspace{-1.5em}
\caption{We present \textsc{Dante-w}, a neural texturing framework for high-fidelity diffuse albedo recovery in the wild. Compared to vanilla mesh texturing with baked-in lighting effects (e.g., noon-time shading and strong roof-edge shadowing on this pavilion), our method effectively disentangles a 3D-consistent diffuse albedo texture with exceptional photorealism. Leveraging physically principled neural rendering, \textsc{Dante-w} faithfully reconstructs fine-grained albedo details, enabling hyper-realistic free-view relighting.}
\label{fig:qualitative_teaser}
\end{figure*}
\end{abstract}    
\section{Introduction}
\label{sec:intro}
Reconstructing high-quality textured assets of real-world large-scale scenes plays a crucial role in domains of virtual reality, filmmaking, and gaming. However, delivering lifelike experiences with these in-the-wild digitizations remains challenging, as it necessitates high-fidelity and 3D-consistent rendering of fine-grained details under arbitrary lighting conditions. In particular, we identify and address the critical limitations of existing methods as follows. 

First of all, existing 3D reconstruction paradigms lack generalizable intrinsic decomposition capability. While traditional image-based 3D reconstruction techniques~\cite{furukawa2015multi, schonberger2016pixelwise, waechter2014let, ling2023large, agisoft} scale well to hundreds of unstructured input images and large-scale scenes, they fail to reason about intrinsic material properties beyond raw pixel observations -- specifically, to disentangle surface reflectance (e.g., diffuse albedo) from illumination effects (i.e., irradiance) -- leading to baked-in shading and shadows that compromise the visual fidelity when varying the illumination. In pursuit of intrinsic decomposition, a wealth of neural inverse rendering approaches~\cite{zhang2021physg, boss2021nerd, Munkberg_2022_CVPR, hasselgren2022shape, jin2023tensoir, li2024idarb, liang2024gs, kaleta2025lumigauss} integrate physically-based rendering~\cite{karis2013real, burley2012physically} with neural representations~\cite{mildenhall2021nerf, wang2021neus, muller2022instant, kerbl20233d}, which jointly reconstruct shape, illumination, and material by imposing learned priors~\cite{zhang2021nerfactor, boss2021neural, litman2025materialfusion, litman2025lightswitch} or empirical regularizations~\cite{yao2022neilf, gao2024relightable, liu2023nero, rudnev2022nerfosr, SOL-NeRF, chang2024fast, gardner2024sky, bai2025gare, feng20252d, liao2025rosgs}. However, they fall short in generalizing to challenging in-the-wild scenarios featuring complex illumination and highly detailed texture, due to oversimplified rendering formulations and the inherent lighting-material ambiguity.

On the other end of the spectrum, recent 2D generation-based methods~\cite{Zeng2024RGBXID, DiffusionRenderer} demonstrate emergent generalizable albedo reasoning capabilities, by inheriting strong priors from pretrained diffusion models~\cite{rombach2022high, blattmann2023stable, agarwal2025cosmos, alhaija2025cosmos} and finetuning on high-quality, task-specific synthetic datasets~\cite{deitke2023objaverse, zhu2022learning, roberts2021hypersim}. However, these methods operate on a per-image or per-video-clip basis, leading to unstable predictions under drastic viewpoint variations while struggling to accurately represent fine details -- primarily due to the lack of spatial awareness, the probabilistic nature of generative modelling, and the information loss within VAE. Moreover, these diffusion-based renderers remain incompatible with modern graphics pipelines and are inefficient for rendering. These limitations underscore the need to bridge these coarse diffusion clues with expressive 3D representations for robust, in-the-wild intrinsic decomposition.

In light of the above observations, we present \textsc{Dante-w}, a neural texturing framework seamlessly integrating with traditional 3D mesh reconstruction pipelines~\cite{furukawa2015multi, schonberger2016pixelwise, agisoft} for high-fidelity diffuse albedo recovery, with an emphasis on challenging in-the-wild scenarios characterized by: 1) large-scale, near-Lambertian scenes with intricate, fine-grained textural details; 2) strong lighting variations involving complex shading and shadowing; 3) highly unstructured viewpoints spanning diverse observation distances. 

To combine the best of both worlds, we fuse generalizable, view-space albedo priors -- obtained from state-of-the-art diffusion models~\cite{Zeng2024RGBXID, DiffusionRenderer} -- into an expressive, 3D-consistent neural texture, thereby aggregating the view-dependent diffusion outputs to suppress flickering hallucinations. It is worth noting that, though, naively lifting per-view prior is prone to over-smoothed results, due to the inaccurate and view-inconsistent diffusion predictions at fine details. 

To address this issue, we propose to learn accurate details from the original images and resolve the ambiguity through the lens of frequency. Our observation is that in-the-wild lighting effects tend to be smoother than fine-grained albedo variations, which is also supported by prior findings~\cite{ramamoorthi2001efficient} that irradiance can be accurately approximated within 1\% error using only nine spherical harmonic coefficients corresponding to the lowest-frequency modes. To this end, we disentangle high-frequency albedo components from irradiance by explicitly differentiating the frequency bands of their respective neural representations, thereby encouraging high-frequency information to flow into the albedo representation when neural rendering the raw imagery. 

Given a mesh reconstructed and parameterized using off-the-shelf tools~\cite{agisoft}, our neural texturing framework operates efficiently -- requiring only a few minutes for large-scale scenes with thousands of unconstrained multi-view images. The resulting neural texture can be directly baked into a standard UV-map and readily imported into any modern graphics engine (e.g., Blender~\cite{blender}) for physically-based relighting, real-time rendering, and flexible editing. 

To comprehensively benchmark our method, we curate a dataset namely \textit{GigaLit}, containing both in-the-wild scenes and synthetic objects, each characterized by complex, high-resolution textures under diverse lighting conditions. Extensive evaluations demonstrate the state-of-the-art performance of \textsc{Dante-w} -- our method outperforms diffusion-based renderers~\cite{Zeng2024RGBXID, DiffusionRenderer} by $+2.63/3.85$$\text{dB}$ in PSNR and achieves $\sim$$25\%$ lower LPIPS for diffuse albedo recovery, and by $+6.13/7.15$$\text{dB}$ in PSNR and over $40\%$ reduction in LPIPS for relighting, underscoring a substantial advancement towards unprecedented robustness and fidelity for in-the-wild relighting applications. To summarize, our main contributions are as follows:
\begin{itemize}
    \item We present \textsc{Dante-w} to recover high-fidelity diffuse albedo for in-the-wild scenes, by lifting view-space diffusion priors to a consistent neural texture representation.

    \item We propose a physically principled neural rendering framework to disentangle 3D-consistent, fine-grained albedo details from the original imagery via a frequency decomposition perspective.

    \item We introduce \textit{GigaLit}, a benchmark dataset for in-the-wild diffuse albedo recovery and relighting, featuring unprecedented textural details, complex illumination variations, and unstructured viewpoints, where our method demonstrates significant improvements over prior works.
\end{itemize}
\section{Related work}
\label{sec:related_work}

\subsection{Classical mesh texturing}
Traditional mesh texturing methods~\cite{waechter2014let, bi2017patch, huang2018gpvc, ling2023large, agisoft} take posed multi-view images and a parameterized mesh from multi-view stereo~\cite{furukawa2015multi, schonberger2016pixelwise} as inputs, and optimize a Markov random field (MRF) energy, where the data term encodes view selection strategies~\cite{waechter2014let, huang2018gpvc, ling2023large} and the smoothness term enforces texture seam consistency via various color adjustment schemes~\cite{lempitsky2007seamless, shen2016color}. Among them, Bi et al.~\cite{bi2017patch} introduces a global patch-based optimization method to enable high-quality, aligned texture mapping from misaligned input images and noisy geometry. Ling et al.~\cite{ling2023large} uses parallelized Loopy Belief Propagation (LBP) to enable per-face view selection, which effectively mitigates non-linear illumination differences with a hierarchical blending scheme. However, these vanilla texturing methods directly operate on pixel colour, thus leaving complex shading and shadowing effects baked into the resulting texture. This limitation substantially degrades visual fidelity in downstream relighting applications. 

\subsection{Intrinsic decomposition}
Intrinsic decomposition aims to separate lighting and underlying surface reflectance from a single image~\cite{bell2014intrinsic, careagaIntrinsic, Zeng2024RGBXID, careaga2024colorful} or multi-view image sequence~\cite{boss2021nerd, DiffusionRenderer, Munkberg_2022_CVPR, jin2023tensoir, li2024idarb}. 
The majority of methods leverage physically-based rendering~\cite{Munkberg_2022_CVPR, hasselgren2022shape, jin2023tensoir, liu2023nero, gao2024relightable, liang2024gs}, generative material estimation~\cite{li2024idarb, wu2025stableintrinsic, litman2025lightswitch}, or a combination of both~\cite{chen2024intrinsicanything, litman2025materialfusion} to jointly recover shape and reflectance for 3D relighting. Some recent works perform diffusion-based generative relighting~\cite{jin2024neural, poirier2024diffusion, zhao2024illuminerf, tangrogr, alzayer2025generative} directly in the image domain without explicit intrinsic reasoning. Despite their effectiveness on small-scale, object-centric scenarios~\cite{kuang2023stanfordorb, deitke2023objaverse} with highly structured capture setting, they fail to generalize to complex, large-scale scenes with unconstrained image collections. Another line of works~\cite{kaleta2025lumigauss, feng20252d, liao2025rosgs} extends inverse rendering to in-the-wild scenarios by imposing tailored priors on sky~\cite{gardner2024sky, bai2025gare} or shadow and sunlight~\cite{rudnev2022nerfosr, chang2024fast, SOL-NeRF, bai2025gare}. However, they struggle to accurately estimate diffuse albedo that is fully disentangled from irradiance, as the imposed priors are insufficient to resolve the inherent lighting-material ambiguity.

On the other hand, recent advances~\cite{Zeng2024RGBXID, DiffusionRenderer} demonstrate compelling generalization capabilities by finetuning general purpose image~\cite{rombach2022high} or video generation models~\cite{agarwal2025cosmos, alhaija2025cosmos} on high-quality synthetic data~\cite{deitke2023objaverse, zhu2022learning, roberts2021hypersim} for in-the-wild intrinsic decomposition. However, their diffusion outputs suffer from the loss of details and 3D-inconsistent hallucinations, thus failing to empower high-fidelity, free-viewpoint rendering. In this work, we propose to aggregate view-space albedo priors from these state-of-the-art diffusion-based renderers into a 3D-consistent neural texture, while leveraging physically-principled neural rendering to further disentangle high-frequency albedo details from the original imagery. Our method effectively eliminates the ambiguity and demonstrates superior expressivity for challenging in-the-wild scenarios. Unlike existing frameworks, we explicitly formulate diffuse albedo recovery within the mesh texturing phase to ensure seamless compatibility with well-established 3D reconstruction and graphics pipelines.

\section{Method}
\label{sec:method}

In this section, we introduce \textsc{Dante-w}, aiming to recover high-fidelity diffuse albedo texture for in-the-wild scenes. Given a collection of unstructured multi-view images and a mesh reconstructed and parameterized using off-the-shelf 3D reconstruction tools, we lift view-dependent, diffusion-based albedo priors to a consistent and expressive 2D neural field defined on the UV-texture space. To enhance fine-grained, 3D-consistent albedo details, we present a physically principled neural rendering framework that disentangles albedo from irradiance by differentiating the frequency bands of their neural representations while jointly optimizing for neural rendering loss and albedo regularization. The resulting neural texture can then be baked into standard texture map and seamlessly integrated with modern graphics engines for relighting.

In the following, we first briefly introduce our shading formulation and neural encoding module (\cref{subsec: prelim}). We then elaborate on our neural texture representation for diffuse albedo (\cref{subsec: albedo}) and our neural rendering framework to recover fine albedo details (\cref{subsec: detail}).

\subsection{Preliminaries}
\label{subsec: prelim}
\par\noindent\textbf{Diffuse shading model.} In this work, we focus on in-the-wild scenes with approximately Lambertian surfaces, consistent with the assumptions commonly adopted in traditional 3D reconstruction literature~\cite{furukawa2015multi, schonberger2016pixelwise, waechter2014let}. Our goal, however, is to take a step further by decomposing the observed colour $\boldsymbol{c} \in \mathbb{R}^3$ into its diffuse albedo $\boldsymbol{a}_d \in \mathbb{R}^3$ and diffuse irradiance $\boldsymbol{s}_d \in \mathbb{R}^3$ components, according to the image formation model:
\begin{equation}
\label{eq:shading}
    \boldsymbol{c} = \boldsymbol{a}_d \odot \boldsymbol{s}_d,
\end{equation}
where the diffuse albedo $\boldsymbol{a}_d$ describes the intrinsic base color of dielectric opaque surfaces, and the diffuse irradiance $\boldsymbol{s}_d$ specifies the incident lighting at a surface position integrated over the upper cosine-weighted hemisphere~\cite{kajiya1986rendering}, which contributes to diverse shading and shadowing effects. Therefore, texturing with $\boldsymbol{a}_d$ instead of $\boldsymbol{c}$ significantly improves visual fidelity under novel lighting conditions. 

\par\noindent\textbf{Multi-resolution hash encoding.}
Recent advances in neural fields~\cite{li2023neuralangelo, barron2023zipnerf, muller2022instant} adopt a hybrid representation that combines multi-resolution hash encoding~\cite{muller2022instant} with a tiny multilayer perceptron neural network (MLP) decoder to faithfully capture fine details. Specifically, hash encoding allocates a hierarchy of independent feature grids across multiple spatial resolutions, where each grid vertex is mapped to a hash entry storing a learnable high-dimensional feature vector. Let $\{V_\ell\}_{\ell=1}^{L}$ be the set of spatial grid resolutions, and given an input position $\boldsymbol{x}$ and a resolution $V_\ell$, we denote by $\boldsymbol{\psi}_{\ell}(\boldsymbol{x}) \in \mathbb{R}^{Z}$ the $Z$-dimensional hash feature obtained via linear interpolation of hash
entries at the vertices of the grid cell enclosing $\boldsymbol{x}$. The hash features across all resolutions are concatenated as: 
\begin{equation}
\label{eq:hash_enc}
    \boldsymbol{\psi}(\boldsymbol{x}) = \big(\boldsymbol{\psi}_{1}(\boldsymbol{x}), ..., \boldsymbol{\psi}_{L}(\boldsymbol{x})\big) \in \mathbb{R}^{LZ},
\end{equation}
which are subsequently fed into the lightweight MLP to decode the desired scene attributes.

\begin{figure*}[htbp]
    \centering
    \newcommand{\colw}{0.19}
    \newcommand{\figw}{1} 
    \includegraphics[width=\figw\textwidth,trim={0cm 0cm 0cm 0cm},clip]{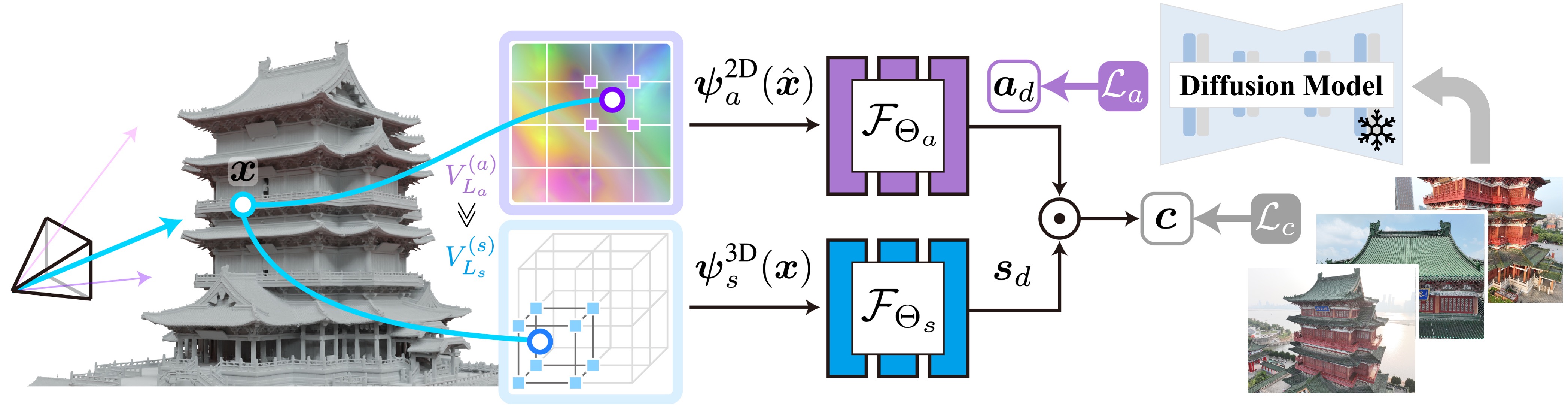}
    \hfill
\vspace{-1.5em}
\caption{An overview of our physically principled neural rendering framework. Given a reconstructed mesh of the scene with surface parameterization, we represent diffuse albedo texture $\boldsymbol{a}_d$ using a high-resolution 2D hash encoding $\boldsymbol{\psi}^{\text{2D}}_a(\cdot)$ and irradiance $\boldsymbol{s}_d$ using a low-resolution 3D hash encoding $\boldsymbol{\psi}^{\text{3D}}_s(\cdot)$. This explicit frequency-band discrepancy (with the maximal grid resolutions satisfying $V_{L_{a}}^{(a)} \gg V_{L_{s}}^{(s)}$) effectively facilitates the disentanglement between the two intrinsic components. We guide the low-frequency component of diffuse albedo with view-space diffusion priors (\cref{eq:albedo_lift}) and recover fine-grained albedo details by neural rendering the raw observations (\cref{eq:loss_color}). 
}
\label{fig:pipeline}
\end{figure*}

\subsection{Albedo representation}
\label{subsec: albedo}
\par
\begin{wrapfigure}{r}{0.4\textwidth}
    \vspace{-0.4cm}
    \centering
    \includegraphics[width=\linewidth]{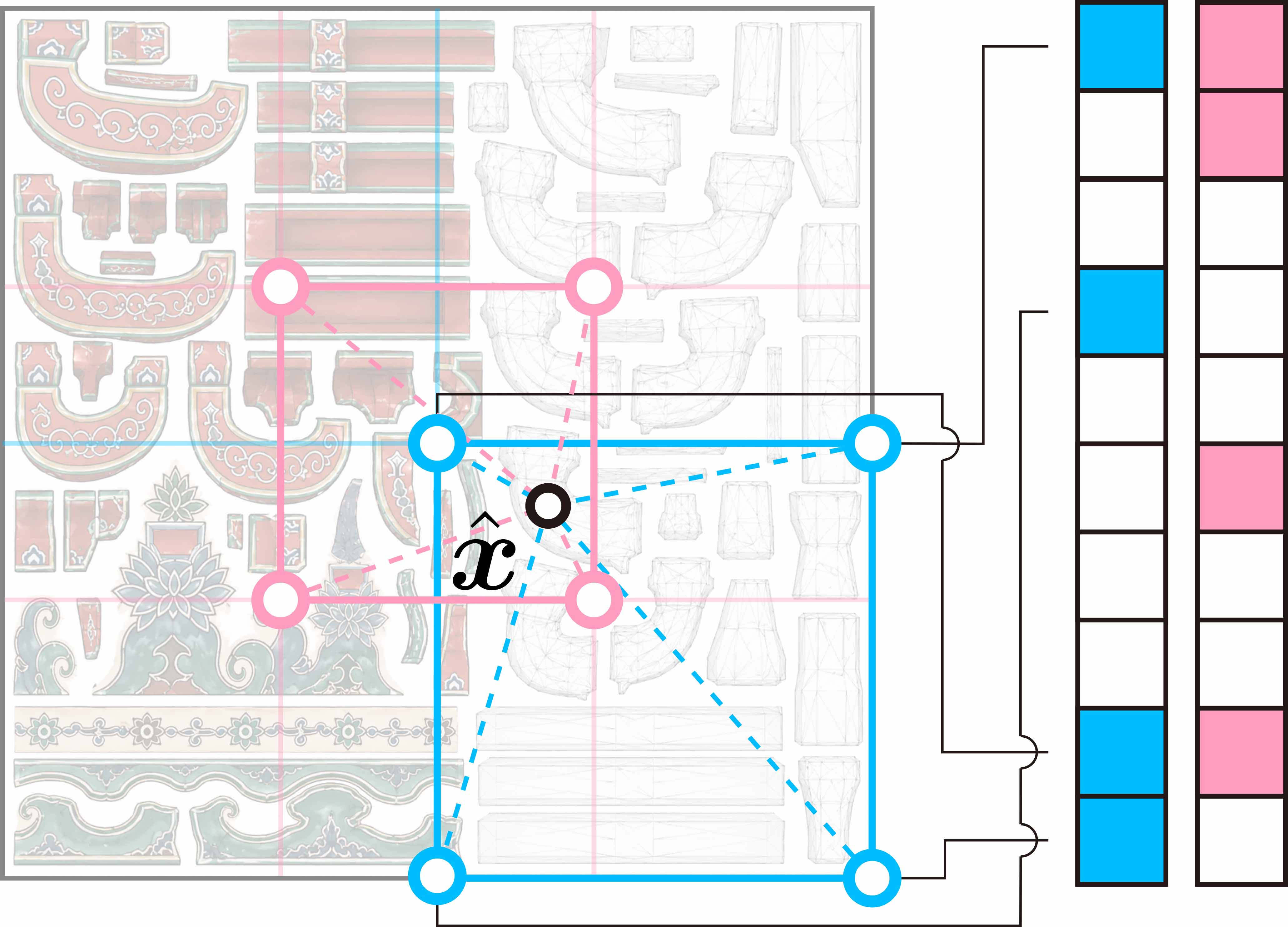}
    \caption{An illustration of the proposed hash-encoded neural texture representation.}
    \vspace{-0.4cm}
\label{fig:representation}
\end{wrapfigure}

\noindent\textbf{Hash-encoded neural texture.} 
Given a reconstructed mesh with its surface parameterization, we represent the diffuse albedo texture in terms of an expressive neural field defined in the UV-coordinate system, as shown in \cref{fig:representation}. 

Specifically, we apply multi-resolution hash encoding~\cite{muller2022instant} in the 2D texture space with a hierarchy of spatial grid resolutions $\{V_\ell^{(a)}\}_{\ell=1}^{L_a}$. We then use a lightweight MLP, $\mathcal{F}_{\Theta_a}$, to interpret the hash features and decode the diffuse albedo value. Let $\hat{\boldsymbol{x}} \in \mathbb{R}^2$ be the parametric UV-coordinate of a surface position, we denote by $\boldsymbol{\psi}^{\text{2D}}_a(\hat{\boldsymbol{x}}) \in \mathbb{R}^{L_aZ_a}$ the concatenated hash features across all $L_a$ resolutions queried at $\hat{\boldsymbol{x}}$, then the diffuse albedo $\boldsymbol{a}_d \in \mathbb{R}^3$ is modelled as:
\begin{equation}
\label{eq:albedo_tex}
    \boldsymbol{a}_d = \mathcal{F}_{\Theta_a}(\boldsymbol{\psi}^{\text{2D}}_a(\hat{\boldsymbol{x}})).
\end{equation}
Benefited from the expressiveness of multi-resolution hash encoding, our neural albedo texture effectively represents high-resolution, intricate details in a memory-efficient manner. Moreover, by organizing a coarse-to-fine hierarchy of spatial grid resolutions, our neural texture inherently mitigates texture seams without the need for any tailored designs~\cite{lempitsky2007seamless}.

\par\noindent\textbf{Lifting view-space albedo priors.} Since disentangling diffuse albedo from raw observations is a highly ill-posed problem, we leverage state-of-the-art intrinsic decomposition diffusion models (e.g., RGB$\leftrightarrow$X~\cite{Zeng2024RGBXID} or DiffusionRenderer~\cite{DiffusionRenderer}) to impose a strong guidance for albedo estimation. To achieve this, we first run the pretrained diffusion model on input images $\{\mathcal{I}_k\}$ to generate the albedo buffer $\{\mathcal{A}^{\prime}_k\}$. We then distill our neural texture using these view-space albedo priors via rasterization and neural rendering:
\begin{equation}
\label{eq:albedo_lift}
    \mathcal{L}_a = \sum_{k}\sum_{\boldsymbol{r}} \|\boldsymbol{a}_d - \mathcal{A}^{\prime}_k(\boldsymbol{r})\|.
\end{equation}
To elaborate, for each pixel ray $\boldsymbol{r}$ from the $k$-th image, we leverage rasterization to obtain the UV-coordinate $\hat{\boldsymbol{x}}$ of the corresponding ray-surface intersection and calculate the diffuse albedo value $\boldsymbol{a}_d$ using \cref{eq:albedo_tex}. We jointly optimize the learnable hash entries and the MLP decoder by minimizing the discrepancies between our albedo texture $\boldsymbol{a}_d$ and the corresponding screen-space albedo value $\mathcal{A}^{\prime}_k(\boldsymbol{r})$ generated by the pretrained diffusion model.

\par\noindent\textbf{Remaining issues.} We find naively lifting view-dependent diffusion outputs (\cref{eq:albedo_lift}) leads to suboptimal results, with the distilled albedo texture exhibiting excessive blurriness, as shown in Fig.~\ref{fig:qualitative_wild}, 
\ref{fig:qualitative_syn_main}, \ref{fig:qualitative_twg}, \ref{fig:supp_orb} (w/o PR). We attribute this degradation to two primary factors: 1) the 2D generative framework fails to maintain 3D consistency, particularly for highly intricate details; 2) the VAE struggles to reconstruct fine-grained details accurately due to the inherent information loss of the compressed latent space. These limitations motivate us to further recover high-fidelity albedo details via physically-based neural rendering.

\subsection{Recovering fine-grained albedo texture}
\label{subsec: detail}
Leveraging diffusion priors as strong guidance for the low-frequency component of diffuse albedo, we propose to further learn 3D consistent, high-frequency albedo details from the raw input images through a physically principled neural rendering framework. Observing that fine albedo details generally exhibit higher spatial frequencies than irradiance for in-the-wild scenes, we can mitigate the albedo-irradiance ambiguity by exploiting their frequency discrepancy within the underlying neural representations.

\par\noindent\textbf{Lighting representation.} Unlike diffuse albedo, which is defined only on the scene surface and can be effectively represented in the UV texture space, light interacts intricately with the scene geometry in 3D space through multiple bounces, yielding shading and shadows shaped by the full set of participating scene structures within the light transport. Therefore, we model global illumination as a volumetric neural field to enable implicit light transport reasoning over the complete 3D scene geometry. 

To elaborate, we leverage hash encoding to represent lighting, with an explicit control over its frequency band by setting the maximal grid resolution. Let $\boldsymbol{x} \in \mathbb{R}^3$ be the 3D coordinate of a surface position, we denote by $\{V_\ell^{(s)}\}_{\ell=1}^{L_s}$ the set of 3D feature grid resolutions and by $\boldsymbol{\psi}^{\text{3D}}_s(\boldsymbol{x}) \in \mathbb{R}^{L_sZ_s}$ the concatenated lighting features across all $L_s$ grid resolutions. To account for unconstrained lighting variations, we introduce a per-image learnable embedding $\boldsymbol{e}_k \in \mathbb{R}^{E}$ and concatenate it with the multi-resolution lighting features~\cite{martin2021nerf, wang2024xscale}. Let $\mathcal{F}_{\Theta_s}$ be a lightweight MLP head, the irradiance $\boldsymbol{s}_d \in \mathbb{R}^3$ is calculated as:
\begin{equation}
\label{eq:lit_tex}
    \boldsymbol{s}_d = \mathcal{F}_{\Theta_s}(\boldsymbol{\psi}^{\text{3D}}_s(\boldsymbol{x}), \boldsymbol{e}_k).
\end{equation}

\par\noindent\textbf{Physically principled neural rendering.} An overview of our full pipeline is illustrated in \cref{fig:pipeline}, where we represent diffuse albedo $\boldsymbol{a}_d$ and irradiance $\boldsymbol{s}_d$ respectively using \cref{eq:albedo_tex} and \cref{eq:lit_tex}, and we compose the final RGB colour according to \cref{eq:shading}. 
Crucially, we disambiguate albedo and irradiance by 1) imposing diffusion-based albedo guidance (\cref{eq:albedo_lift}) and 2) differentiating their maximal grid resolutions, i.e., $V_{L_{a}}^{(a)} \gg V_{L_{s}}^{(s)}$. This design enables fine-grained details to flow into the diffuse albedo component when recreating the raw input image:
\begin{equation}
\label{eq:loss_color}
    \mathcal{L}_c = \sum_{k}\sum_{\boldsymbol{r}} \|\gamma(\boldsymbol{c}) - \mathcal{I}_k(\boldsymbol{r})\|,
\end{equation}
where we denote by $\mathcal{I}_k(\boldsymbol{r})$ the raw RGB value at pixel location $\boldsymbol{r}$ of the $k$-th input image, and by $\gamma(\cdot)$ a fixed tone mapping function that converts linear color to
sRGB space~\cite{anderson1996proposal}. 



Our full loss function involves the neural rendering loss (\cref{eq:loss_color}) and the diffusion-based albedo regularization term (\cref{eq:albedo_lift}), which are jointly optimized with respect to the learned hash entries and the MLP parameters: 
\begin{equation}
\label{eq:loss}
    \mathcal{L} = \lambda \mathcal{L}_a + \mathcal{L}_c,
\end{equation}
where $\lambda$ denotes the relative weight of the albedo regularization and we empirically set it to $1.0$ for all scenes in our implementation.

\subsection{Implementation details}

\par\noindent\textbf{Exposure compensation.} To compensate for unconstrained camera response, we introduce a pair of learnable scalars $w_k, b_k \in \mathbb{R}$ for each input image $\mathcal{I}_k$ to rectify the computed RGB color $\boldsymbol{c}$~\cite{chen2024pgsr}.

\par\noindent\textbf{Texture map baking.} Our hash-encoded neural texture can be readily transformed to standard texture map, making our framework integrate seamlessly with well-established graphics pipelines. To generate the texture map, we first initialize a UV-map at the desired texture resolution, with each pixel storing its 2D UV-coordinate $\hat{\boldsymbol{x}}$. This UV-map is then passed through our neural albedo renderer (\cref{eq:albedo_tex}) to produce the diffuse albedo map. 

\par\noindent\textbf{Ray batching.} We cache the rasterization buffer for all input views prior to training, which stores, for each pixel, the 3D and UV coordinates of the ray-surface intersection and a binary foreground indicator. During training, we randomly sample a set of foreground pixel rays across all views for each gradient descent step. We empirically find this strategy leads to slightly better performance and faster convergence.
\section{Experiments}
\label{sec:exp}

\subsection{GigaLit benchmark}
Existing benchmarks~\cite{kuang2023stanfordorb, jin2023tensoir, deitke2023objaverse} for intrinsic decomposition and relighting are limited to highly constrained, object-centric scenarios characterized by: 1) simple albedo textures with minimal fine-grained details; 2) static illumination represented by an environment map lacking complex variations in shading and shadowing; 3) evenly distributed viewpoints centered around the object at a fixed observation scale. 

In light of these limitations, we introduce \textit{GigaLit}, a new benchmark dataset designed to characterize the complexity and richness of in-the-wild scenes for diffuse albedo recovery and relighting. Specifically, GigaLit contains $6$ real-world outdoor scenes for qualitative evaluation, featuring intricate details, strong shading, and complex shadowing effects. For each scene, we capture hundreds to thousands of unstructured, high-quality images across multiple observation distances and different times of the day using a DSLR camera. We then run Metashape Pro~\cite{agisoft} for structure-from-motion, multi-view stereo, surface reconstruction and parameterization. 

To support quantitative evaluation, we also meticulously curate $10$ synthetic objects with fine albedo details and 4K resolution texture from BlenderKit~\cite{blenderkit}. We leverage Blender~\cite{blender} to simulate challenging lighting variations by altering environment maps collected from PolyHeaven~\cite{polyheaven} and directional spot lights. To mimic the unstructured nature of in-the-wild data, we render each synthetic object along an upper hemispherical trajectory with varying camera-to-object distances. We refer readers to the supplement for additional dataset details.

\par\noindent\textbf{Baselines.} We test our method with two state-of-the-art intrinsic decomposition models -- RGB$\leftrightarrow$X~\cite{Zeng2024RGBXID} and Cosmos-DiffusionRenderer (DR)~\cite{DiffusionRenderer} -- both supporting forward and inverse rendering via diffusion-based inference. We compare against their raw diffusion outputs, vanilla mesh texturing methods~\cite{waechter2014let, bi2017patch, agisoft}, and LightSwitch~\cite{litman2025lightswitch}, which performs material estimation and generative relighting via multi-view diffusion. We also compare with the ablated variants of our framework without physically principled neural rendering ([Ours+RGB$\leftrightarrow$X/DR] (w/o PR)) throughout all experiments. For mesh-based methods, we represent diffuse albedo as standard UV texture map and perform relighting using Blender’s Cycles engine~\cite{iraci2013blender}.

\par\noindent\textbf{Experimental setup.} For real-captured, in-the-wild scenes, we qualitatively compare all baselines by recovering diffuse albedo texture under the varied illuminations during data collection and relighting with novel environment maps from PolyHeaven. For synthetic data, we pair each object with $5$ different environment maps to simulate the unconstrained lighting for albedo recovery, where each input view of the object is rendered using a randomly selected one. We hold out another $5$ environment maps shared across objects for relighting evaluation. We use PSNR, SSIM and LPIPS as quantitative metrics, and report the performance as the average over 10 objects, 250 views, and 5 lighting conditions. 

\begin{figure*}[htbp]
    \centering
    \newcommand{\colw}{0.2}
    \newcommand{\figw}{1.0} 
    \includegraphics[width=\figw\textwidth,trim={0cm 0cm 0cm 0cm},clip]{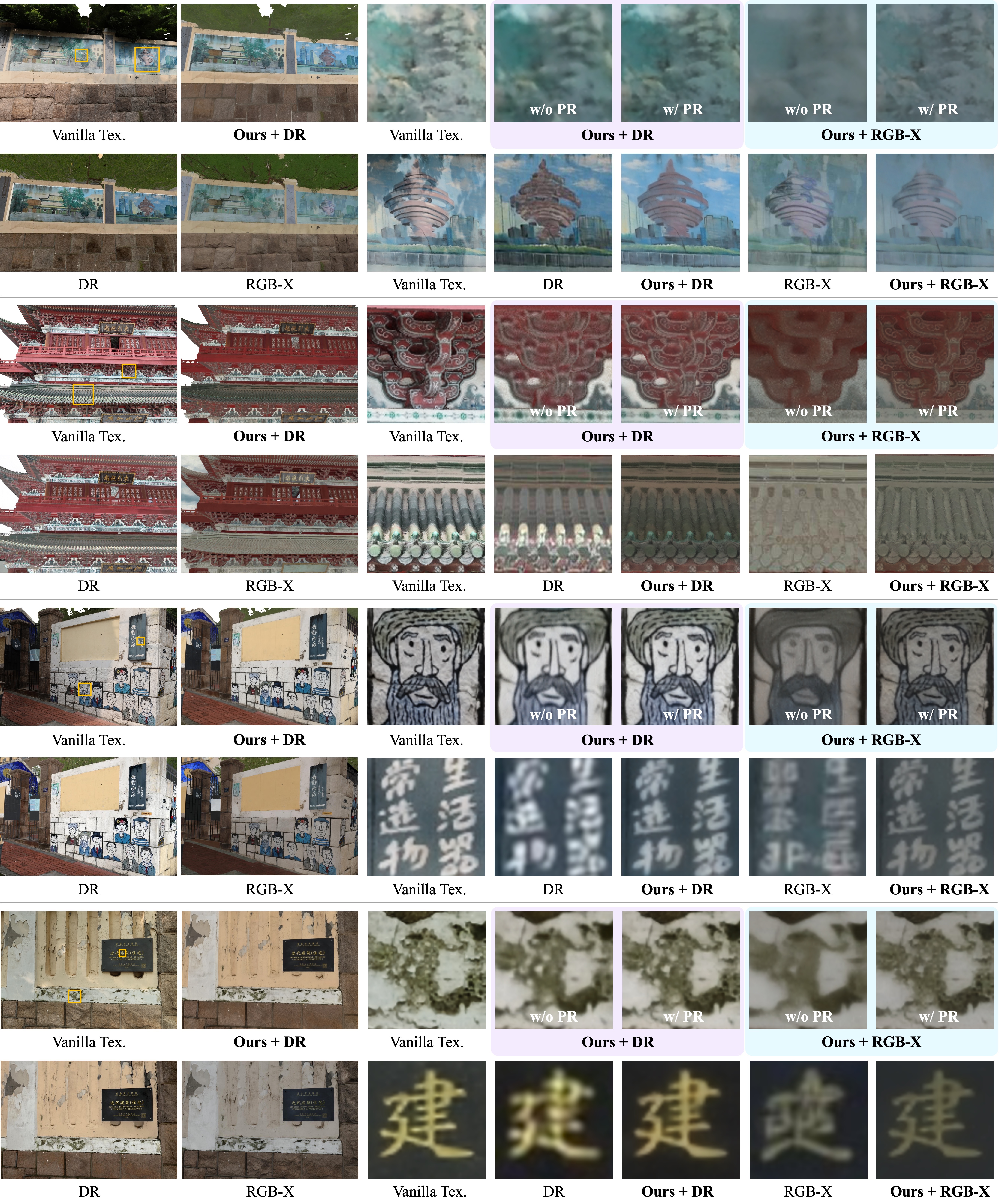}
\hfill
\vspace{-2.0em}
\caption{Visual comparison of diffuse albedo recovery on in-the-wild scenes. By lifting view-space diffusion priors upon a coherent neural texture via physically principled neural rendering (PR), our method robustly eliminates baked-in shading and shadowing effects that remain challenging for vanilla mesh texturing (Metashape Pro~\cite{agisoft}), while refining raw diffusion outputs of intrinsic decomposition models (RGB$\leftrightarrow$X~\cite{Zeng2024RGBXID} and Cosmos-DiffusionRenderer (DR)~\cite{DiffusionRenderer}) with faithful, fine-grained, 3D-consistent details.}
\label{fig:qualitative_wild}
\end{figure*}
\begin{figure*}[t]
    \centering
    \newcommand{\colw}{0.2}
    \newcommand{\figw}{1.0} 
    \includegraphics[width=\figw\textwidth,trim={0cm 0cm 0cm 0cm},clip]{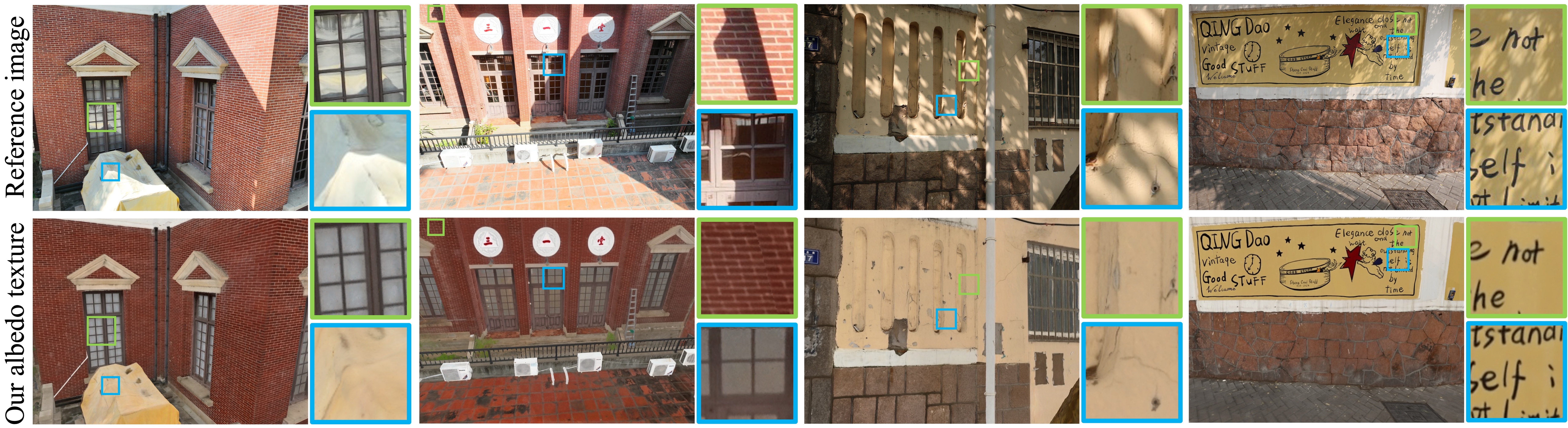}
\hfill
\vspace{-2.0em}
\caption{Generalization on sharp shadows and strong specularity using GigaLit scenes.}
\label{fig:qualitative_hard}
\end{figure*}

\subsection{Comparative results}
\par\noindent\textbf{Results on in-the-wild scenes.} The visual comparisons on GigaLit in-the-wild scenes are presented in \cref{fig:qualitative_wild}. As observed, vanilla mesh texturing methods like Metashape Pro~\cite{agisoft} fail to disentangle shading and shadowing from the underlying diffuse albedo, leaving these irradiance components baked into the UV texture. For recent diffusion-based renderers such as Cosmos-DiffusionRenderer (DR)~\cite{DiffusionRenderer} and RGB$\leftrightarrow$X~\cite{Zeng2024RGBXID}, their raw diffusion outputs primarily lack accuracy, stability and 3D consistency across varying viewpoints, particularly at fine details. By contrast, our method enables strictly 3D-consistent and more faithful diffuse albedo recovery. As shown in \cref{fig:qualitative_hard}, our method also generalizes well to more complex scenarios, e.g., sharp shadows and strong reflections, due to the robustness of meshing and our 3D-consistent neural texture. We refer readers to the supplement for more qualitative results.

\par
\begin{figure*}[htbp]
    \centering
    \newcommand{\colw}{0.2}
    \newcommand{\figw}{1.0} 
    \includegraphics[width=\figw\textwidth,trim={0cm 0cm 0cm 0cm},clip]{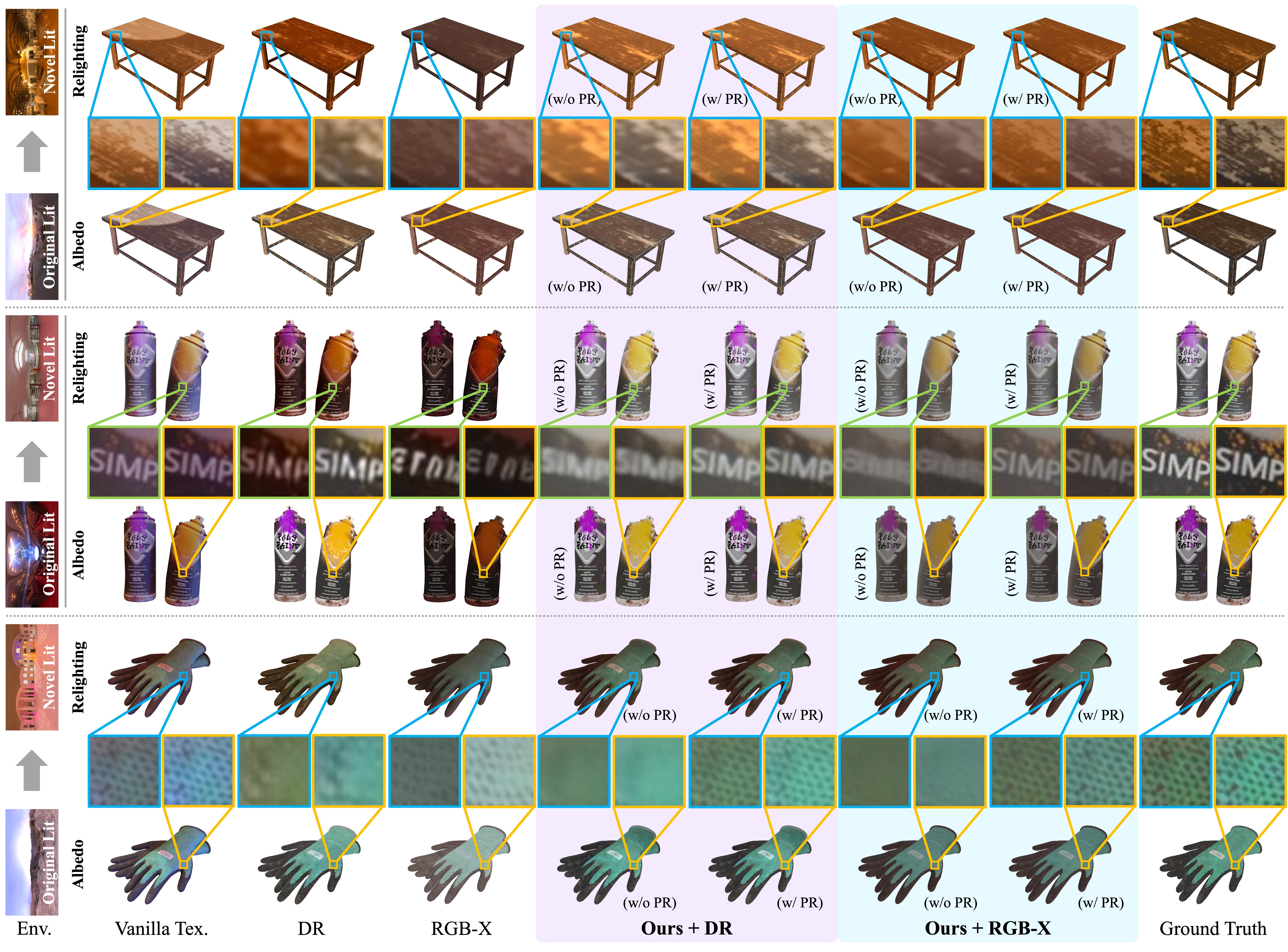}
\hfill
\vspace{-2.0em}
\caption{Qualitative results of diffuse albedo recovery and relighting on GigaLit synthetic objects, 
compared with vanilla texturing (Metashape Pro~\cite{agisoft}), raw outputs of RGB$\leftrightarrow$X~\cite{Zeng2024RGBXID} and Cosmos-DiffusionRenderer (DR)~\cite{DiffusionRenderer}, and the ablated variant of our method in terms of physically principled neural rendering (PR). The albedo is recovered under the \textit{Original Lit} and subsequently relighted under the \textit{Novel Lit}. 
Our method effectively removes strong lighting effects and faithfully preserves fine details, while achieving superior relighting fidelity by capitalizing on the standard UV texture and physically-based ray tracing~\cite{iraci2013blender}.
}
\label{fig:qualitative_syn_main}
\end{figure*}

\noindent\textbf{Results on synthetic data.} In \cref{tab:quantitative}, we report the mean metrics on the GigaLit synthetic subset. Our method significantly outperforms all baselines in terms of albedo recovery, yielding $+4.49\text{dB}$ PSNR over Metashape Pro~\cite{agisoft}, $+3.85\text{dB}$ PSNR and $\sim$$25\%$ reduction in LPIPS relative to RGB$\leftrightarrow$X~\cite{Zeng2024RGBXID}, and $+2.63\text{dB}$ PSNR and $\sim$$24\%$ lower LPIPS relative to Cosmos-DiffusionRenderer (DR)~\cite{DiffusionRenderer}. For the task of relighting, our method also demonstrate strong performance, achieving $+3.94\text{dB}$ PSNR over Metashape Pro~\cite{agisoft}, $+7.15\text{dB}$ PSNR and $\sim$$43\%$ lower LPIPS relative to RGB$\leftrightarrow$X, and $+6.13\text{dB}$ PSNR and $\sim$$45\%$ reduction in LPIPS relative to DiffusionRenderer. Note that LightSwitch~\cite{litman2025lightswitch} struggles to generalize to complex, varied illumination and unstructured viewpoints, yielding significant hallucinations in both albedo and relighting. The qualitative comparisons are presented in \cref{fig:qualitative_syn_main}. Under diverse, strong lighting conditions, diffusion-based renderers~\cite{Zeng2024RGBXID, DiffusionRenderer} struggle in physically-accurate relighting, whereas our method demonstrates exceptional robustness by recovering faithful and detailed albedo texture and relighting with physically-based ray tracing~\cite{iraci2013blender} and well-established graphics engines~\cite{blender}.

\begin{table}
\caption{Quantitative evaluations on GigaLit synthetic dataset.}
\label{tab:quantitative}
\centering
\resizebox{\linewidth}{!}{
\newcommand{\tabincell}[2]{\begin{tabular}{@{}#1@{}}#2\end{tabular}}
\setlength{\tabcolsep}{2.0mm}{
\begin{tabular}{c|ccc|ccc}
\toprule
\multirow{2}{*}{\tabincell{c}{}} & \multicolumn{3}{c|}{Diffuse Albedo} & \multicolumn{3}{c}{Relighting} \\ 
{} & PSNR$\uparrow$ & SSIM$\uparrow$ & LPIPS$\downarrow$ & PSNR$\uparrow$ & SSIM$\uparrow$ & LPIPS$\downarrow$\\ \midrule
Waechter et al.~\cite{waechter2014let} & 21.43 & 0.889 & 0.103 & 23.78 & 0.915 & 0.092 \\
Bi et al.~\cite{bi2017patch} & 21.48 & 0.885 & 0.111 & 23.81 & 0.902 & 0.097 \\
Metashape Pro~\cite{agisoft} & 21.50 & 0.892 & 0.103 & 23.83 & 0.917 & 0.089 \\
\midrule
LightSwitch~\cite{litman2025lightswitch} & 16.89 & 0.807 & 0.185 & 17.79 & 0.808 & 0.189 \\
\cellcolor{x3}RGB$\leftrightarrow$X~\cite{Zeng2024RGBXID} & \cellcolor{x3}21.65 & \cellcolor{x3}0.860 & \cellcolor{x3}0.125 & \cellcolor{x3}20.30 & \cellcolor{x3}0.849 & \cellcolor{x3}0.134 \\
\cellcolor{dr3}DiffusionRenderer (DR)~\cite{DiffusionRenderer} & \cellcolor{dr3}23.36 & \cellcolor{dr3}0.879 & \cellcolor{dr3}0.120 & \cellcolor{dr3}21.64 & \cellcolor{dr3}0.873 & \cellcolor{dr3}0.130 \\
\midrule
\cellcolor{x3}$[$Ours+RGB$\leftrightarrow$X$]$ (w/o PR) & \cellcolor{x3}25.32 & \cellcolor{x3}0.890 & \cellcolor{x3}0.120 & \cellcolor{x3}27.27 & \cellcolor{x3}0.918 & \cellcolor{x3}0.099 \\ 
\cellcolor{x3}\textbf{$[$Ours+RGB$\leftrightarrow$X$]$ (Full)} & \cellcolor{x3}{25.50} & \cellcolor{x3}{0.911} & \cellcolor{x3}{0.093} & \cellcolor{x3}{27.45} & \cellcolor{x3}{0.933} & \cellcolor{x3}{0.076} \\ 
\cellcolor{dr3}$[$Ours+DR$]$ (w/o PR) & \cellcolor{dr3}25.69 & \cellcolor{dr3}0.907 & \cellcolor{dr3}0.115 & \cellcolor{dr3}27.45 & \cellcolor{dr3}0.928 & \cellcolor{dr3}0.092 \\
\cellcolor{dr3}\textbf{$[$Ours+DR$]$ (Full)} & \cellcolor{dr3}\textbf{25.99} & \cellcolor{dr3}\textbf{0.921} & \cellcolor{dr3}\textbf{0.091} & \cellcolor{dr3}\textbf{27.77} & \cellcolor{dr3}\textbf{0.938} & \cellcolor{dr3}\textbf{0.071} \\
\bottomrule
\end{tabular}
}
}
\end{table}
\begin{figure*}[htbp]
    \centering
    \newcommand{\colw}{0.2}
    \newcommand{\figw}{1.0} 
    \includegraphics[width=\figw\textwidth,trim={0cm 0cm 0cm 0cm},clip]{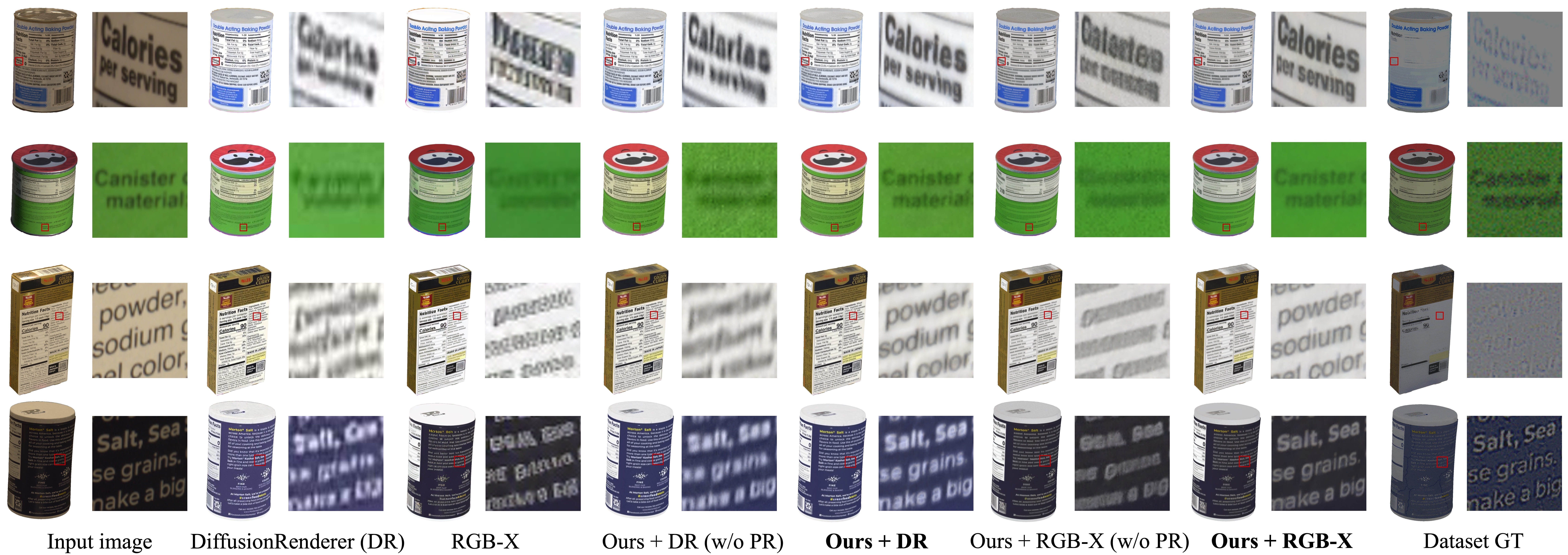}
\hfill
\vspace{-1.5em}
\caption{Qualitative comparisons on diffuse albedo recovery using Stanford-ORB dataset~\cite{kuang2023stanfordorb}. The raw diffusion outputs from both DiffusionRenderer (DR)~\cite{DiffusionRenderer} and RGB$\leftrightarrow$X~\cite{Zeng2024RGBXID} tend to exhibit hallucinated artifacts caused by the probabilistic nature of generative modelling and the information loss of the VAE latent. Naively lifting view-space priors without our physically principled neural rendering (Ours+DR/RGB$\leftrightarrow$X (w/o PR)) also leads to severe blurriness, since the diffusion model fails to reason about 3D consistency and yields view-dependent flickering artifacts. By contrast, our method (Ours+DR/RGB$\leftrightarrow$X) accurately recovers finer details while robustly eliminates strong irradiance effects. Note that the albedo ground truth in Stanford-ORB (Dataset GT) contains noisy capture artifacts that significantly deteriorate fine-grained textures, making it unsuitable for quantitative benchmarking of high-fidelity albedo recovery. Please zoom in to see our high-quality details.}
\label{fig:supp_orb}
\end{figure*}
\begin{figure*}[htbp]
    \centering
    \newcommand{\colw}{0.2}
    \newcommand{\figw}{1.0} 
    \includegraphics[width=\figw\textwidth,trim={0cm 0cm 0cm 0cm},clip]{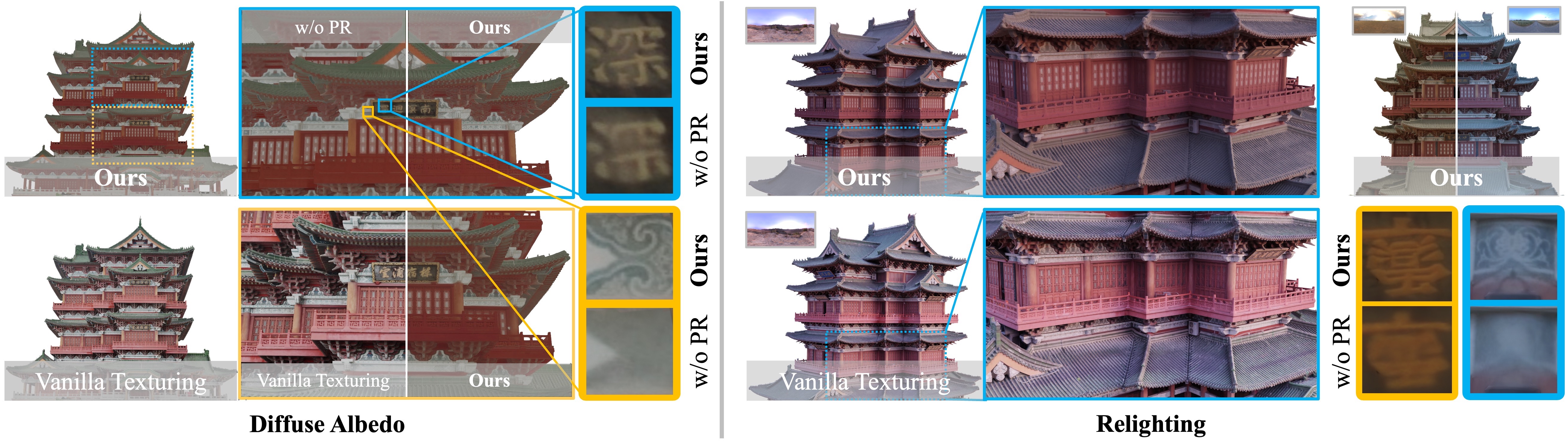}
\hfill
\vspace{-2.0em}
\caption{Visualizations of diffuse albedo recovery and relighting on a large-scale, in-the-wild scene -- The Pavilion of Prince Teng -- reconstructed using over 2,500 unstructured photographs. Our method learns a consistent diffuse albedo texture with significantly finer details compared to the naive lifting of diffusion priors (w/o PR). We also demonstrate superior fidelity compared to the vanilla texturing approach, Metashape Pro~\cite{agisoft}, which is prone to baked-in irradiance. Please zoom in to see details and lighting effects.}
\label{fig:qualitative_twg}
\end{figure*}
\begin{figure*}[htbp]
    \centering
    \newcommand{\colw}{0.2}
    \newcommand{\figw}{1.0} 
    \includegraphics[width=\figw\textwidth,trim={0cm 0cm 0cm 0cm},clip]{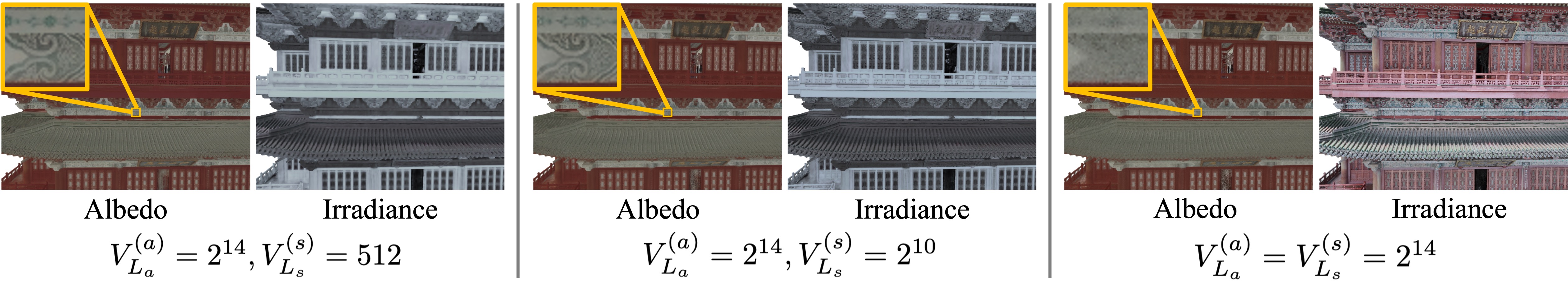}
\hfill
\vspace{-2.0em}
\caption{Ablation study on frequency discrepancy. Differentiating the maximal grid resolution facilitates the disentanglement.
}
\label{fig:ablation_freq}
\end{figure*}
\begin{figure*}[htbp]
    \newcommand{\colw}{0.2}
    \newcommand{\figw}{0.8} 
    \centering
    \includegraphics[width=\figw\textwidth,trim={0cm 0cm 0cm 0cm},clip]
    {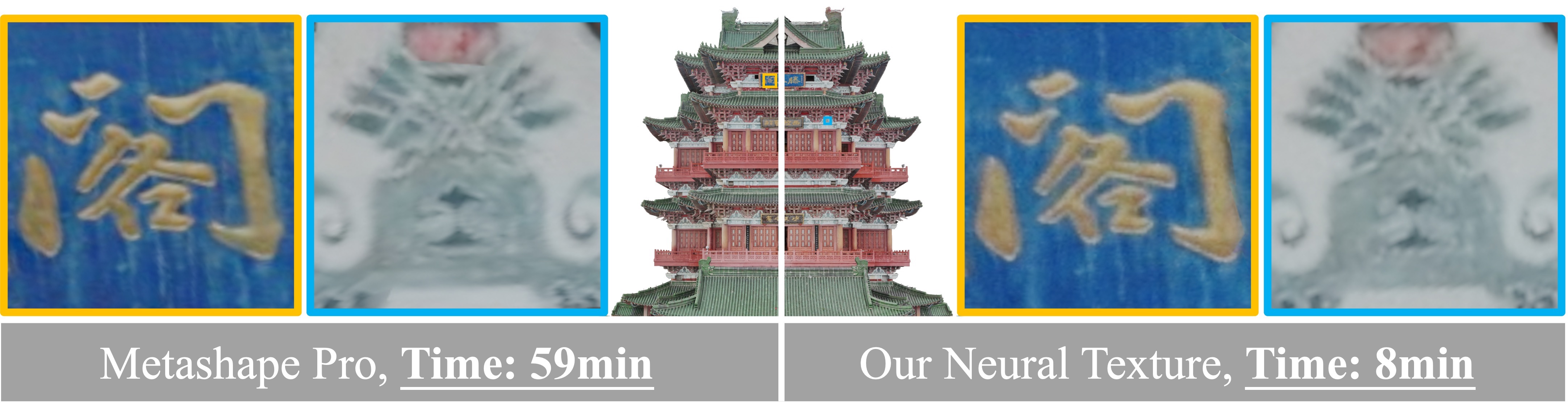}
\hfill
\centering
\caption{Run time comparison with Metashape Pro~\cite{agisoft} in terms of mesh texturing. Our method demonstrates comparable visual quality while achieving over 7$\times$ acceleration. }
\label{fig:tex_compare}
\end{figure*}

\par\noindent\textbf{Results on Stanford-ORB dataset.} We also conduct experiments on the public Stanford-ORB dataset~\cite{kuang2023stanfordorb}, and the visual results of diffuse albedo recovery are presented in \cref{fig:supp_orb}. Compared to raw diffusion outputs from DiffusionRenderer (DR)~\cite{DiffusionRenderer} and RGB$\leftrightarrow$X~\cite{Zeng2024RGBXID}, our method demonstrates significant improvements in recovering 3D-consistent, fine-grained details while maintaining the robustness in irradiance removal -- zoom in to see our high-quality details of the rendered text on the object surface. We also remark that the ground-truth albedo annotations provided by the Stanford-ORB dataset suffer from low capture quality and significant approximation errors, which exhibits pronounced noise and considerable degradation of fine textural details -- see the last column `Dataset GT' of \cref{fig:supp_orb}. Since there are no precise diffuse albedo measurements for real-world scenes, the ground-truth albedo in Stanford-ORB serves only as a coarse reference, failing to provide a principled and rigorous quantitative evaluation for the task of high-fidelity diffuse albedo recovery, particularly for highly intricate textural details. By contrast, our method effectively learns accurate, fine-grained details that exhibit comparable visual fidelity to those in the original input images, significantly outperforming the quality of the ground truth annotations. This limitation motivates us to curate high-quality synthetic data in our GigaLit dataset for more reasonable quantitative evaluations.

\par\noindent\textbf{Ablation studies.}
We ablate the proposed physically principled neural rendering (PR) framework and the comparative results (denoted as (w/o PR)) are shown in Fig.~\ref{fig:qualitative_wild}, 
\ref{fig:qualitative_syn_main}, \ref{fig:supp_orb}, \ref{fig:qualitative_twg}, and \cref{tab:quantitative}. This design effectively compensates for the loss of high-frequency albedo details when lifting inconsistent screen-space albedo priors. We also ablate the frequency discrepancy design in terms of our albedo and lighting representation, and the visual results are presented in \cref{fig:ablation_freq}. Without explicitly differentiating the frequency bands of the underlying neural representations, the neural rendering paradigm suffers from ambiguity that tends to attribute fine details to the irradiance component.

\par\noindent\textbf{Efficiency report.} Leveraging advanced neural encoding implementations~\cite{tiny-cuda-nn}, our diffuse albedo neural texturing framework is highly efficient, with the entire optimization generally taking $5$$\sim$$8$ minutes per scene on a single Nvidia RTX 3090 GPU -- over $7\times$ faster than the multi-view mesh texturing process of Metashape Pro~\cite{agisoft}, while maintaining the comparable visual quality, as shown in \cref{fig:tex_compare}. In terms of storage, our method consumes 0.8GB of GPU memory for object-centric synthetic scenes (maximal hash grid resolution $V_{L_{a}}^{(a)}=2048$), and 3.3GB for in-the-wild scenes ($V_{L_{a}}^{(a)}=8192$).
\section{Conclusion}
\label{sec:conclusion}

We introduce \textsc{Dante-w}, the first high-fidelity diffuse albedo neural texturing approach for complex, in-the-wild scenes, taking as input unconstrained image collections. The core idea is to lift inconsistent view-space diffusion priors into a coherent and expressive neural texture representation, while incentivizing 3D-consistent, highly-intricate albedo details by exploiting frequency-band discrepancy within a physically principled neural rendering framework. 
Extensive experiments demonstrate the superiority of our method in robustly alleviating strong irradiance effects and achieving unparalleled relighting fidelity.

\par\noindent\textbf{Limitation \& Future Work.} 
Despite the effectiveness in lifting view-dependent, diffusion-based albedo priors with accurate, 3D-consistent details, our current method is designed as a mesh texturing step within traditional 3D reconstruction pipeline and therefore does not account for non-Lambertian effects. One future research direction is to exploit expressive neural texture in advanced scene representations (e.g., Gaussian Splatting~\cite{kerbl20233d}) and include specular materials into the proposed framework.

\section*{Acknowledgements}
This work is supported in part by Natural Science Foundation of China (NSFC) under contract No. 62125106 and 62427804, in part by Tsinghua University Dushi Program (No.20251080107), in part by the Beijing Outstanding Young Scientist Program under contract No. JWZQ20240101009, in part by the XPLORER PRIZE.

%
%
\bibliographystyle{splncs04}
\bibliography{main}
\end{document}